\documentclass[aps,pre,reprint]{revtex4-1}
\usepackage{times}

\usepackage{graphicx}
\graphicspath{
  {./pics/}
  {./pics/sublattice_coding/},
  {./pics/parallel/},
  {./pics/benchmarks/},
  {./pics/scaling/},
}
\usepackage{amsmath}
\usepackage{amssymb}
\usepackage{bm}
\usepackage{color}

\usepackage{algorithm}
\usepackage[noend]{algpseudocode}
\algblock[ForEach]{ForEach}{EndForEach}
\algblockdefx[ForEach]{ForEach}{EndForEach}[1]{\textbf{for each } #1 \textbf{:}}{}
\makeatletter
\ifthenelse{\equal{\ALG@noend}{t}}{\algtext*{EndForEach}}{}
\makeatother
\usepackage{caption}
\usepackage{subcaption}
\usepackage[colorlinks=true, urlcolor=blue, linkcolor=red, citecolor=blue, pdftex]{hyperref}
\usepackage{cleveref}
\crefname{equation}{Eq.}{Eqs.}
\Crefname{equation}{Eq.}{Eqs.}
\crefname{figure}{Fig.}{Figs.}
\Crefname{figure}{Fig.}{Figs.}
\crefname{part}{part}{parts}
\Crefname{part}{Part}{Parts}
\crefname{chapter}{chapter}{chapters}
\Crefname{chapter}{Chapter}{Chapters}
\crefname{section}{section}{sections}
\Crefname{section}{Section}{Sections}
\crefname{subsection}{section}{sections}
\Crefname{subsection}{Section}{Sections}
\crefname{appendix}{appendix}{appendices}
\Crefname{appendix}{Appendix}{Appendices}

\DeclareMathOperator{\integer}{int}

\newcommand{\argmin}{\operatornamewithlimits{argmin}}

\usepackage{braket}

\makeatletter
\def\moverlay{\mathpalette\mov@rlay}
\def\mov@rlay#1#2{\leavevmode\vtop{%
   \baselineskip\z@skip \lineskiplimit-\maxdimen
   \ialign{\hfil$\m@th#1##$\hfil\cr#2\crcr}}}
\newcommand{\charfusion}[3][\mathord]{
    #1{\ifx#1\mathop\vphantom{#2}\fi
        \mathpalette\mov@rlay{#2\cr#3}
      }
    \ifx#1\mathop\expandafter\displaylimits\fi}
\makeatother

\newcommand{\bigcupdot}{\charfusion[\mathop]{\bigcup}{\cdot}}

\begin{document}
\title{Sublattice Coding Algorithm and Distributed Memory
  Parallelization for Large-Scale Exact Diagonalizations of Quantum
  Many-Body Systems} \author{Alexander Wietek}
\email{alexander.wietek@gmail.com} \author{Andreas M. L\"auchli}
\affiliation{Institut f\"ur Theoretische Physik, Universit\"at
  Innsbruck, A-6020 Innsbruck, Austria}

\begin{abstract}
  We present algorithmic improvements for fast and memory-efficient
  use of discrete spatial symmetries in Exact Diagonalization
  computations of quantum many-body systems. These techniques allow us
  to work flexibly in the reduced basis of symmetry-adapted wave
  functions. Moreover, a parallelization scheme for the
  Hamiltonian-vector multiplication in the Lanczos procedure for
  distributed memory machines avoiding load balancing problems is
  proposed. We demonstrate that using these methods low-energy
  properties of systems of up to $50$ spin-$1/2$ particles can be
  successfully determined.
\end{abstract}

\maketitle
%%%%%%%%%%%%%%%%%%%%%%%%%%%%%%%%%%%%%%%%%%%%
\section{Introduction}
Exact Diagonalization, short ED, studies have in the past been a
reliable source of numerical insight into various problems in quantum
many-body physics, ranging from quantum
chemistry~\cite{Zhengting2005}, nuclear structure\cite{Sternberg2008,
  Vary2009, Maris2010}, quantum field theory~\cite{Rychkov2015} to
strongly correlated lattice models in condensed matter physics. The
method is versatile, unbiased and capable of simulating systems with a
sign problem. The main limitation of ED is the typically exponential
scaling of computational effort and memory requirements in the system
size. Nevertheless, the number of particles or lattice sites feasible
for simulation has steadily increased since the early
beginnings~\cite{Oitmaa1978} and have provided valuable insight to
many problems in modern condensed matter physics, for example,
frustrated magnetism~\cite{Wietek2017a, Wietek2015,
  Dagotto1989,Leung1993,Schulz1996,Richter2010,Laeuchli2012,Lauchli2016},
high temperature
superconductivity~\cite{Lin1987,Hasegawa1989,Bonca1989,Poilblanc1995},
quantum hall effect and fractional Chern
insulators~\cite{Halperin1994,Morf2002,Neupert2011,Laeuchli2013} and
quantum critical points in 2+1
dimensions~\cite{Schuler2016,Whitsitt2017}.  Different approaches for
increasing the system size in these simulations have been proposed
over time~\cite{Lin1990, Weisse2013}. Not only does increasing the
number of particles yield better approximations to the thermodynamic
limit, but also several interesting simulation clusters with many
symmetries become available if more particles can be simulated. Having
access to such clusters becomes important if several competing phases
ought to be realized on the same finite size sample.

The ED method is essentially equivalent to simulating quantum
circuits.  With the advent of scalable experimental quantum
computation~\cite{Feynman1982,Haffner2005,Castelvecchi2017}, exact
classical simulation of quantum circuits has become important for
benchmarking and validating results from actual quantum
computers~\cite{Haner2017, Pednault2017, Gheorgiu2017}. At present, we
are on the verge of quantum computers surpassing the capabilities of
classical supercomputers in terms of the number of simulated Qubits,
colloquially referred to as quantum advantage. Specifically, the
barrier of classically simulating $50$ Qubits has not been breached to
date.

In this work, we present algorithms and strategies for the
implementation of a state-of-the-art large-scale ED code and prove
that applying these methods systems of up to $50$ spin-$1/2$ particles
can be simulated on present day supercomputers. There are two key
ingredients making these computations possible:
\begin{itemize}
\item \textbf{Efficient use of symmetries}. We present an algorithm to
  work with symmetry-adapted wave functions in a fast and memory
  efficient way. This so-called \textit{sublattice coding algorithm}
  allows us to diagonalize the Hamiltonian in every irreducible
  representation of a discrete symmetry group. The basic idea behind
  this algorithm goes back to H.Q. Lin~\cite{Lin1990}. An extension of
  this method was proposed in Ref.~\cite{Weisse2013}. We generalize
  these approaches to arbitrary discrete symmetries, varying number of
  sublattices and arbitrary geometries.
\item \textbf{Parallelization of the matrix-vector multiplications} in
  the Lanczos algorithm~\cite{Lanczos1950} for distributed memory
  machines. We propose a method avoiding load-balancing problems in
  message-passing and present a computationally fast way of storing
  the Hilbert space basis.
\end{itemize}
These ideas have been implemented and tested on various
supercomputers.  We present results and benchmarks to demonstrate the
efficiency and flexibility of the proposed methods.

\section{Symmetry adapted basis states}

%%%%%%%%%%%%%%%%%%%%%%%%%%%%%%%%%%%%%%%%%%%%
\label{sec:sublatticecodingtechnique}
\begin{figure*}[t!]
  \centering \includegraphics[width=0.8\textwidth]{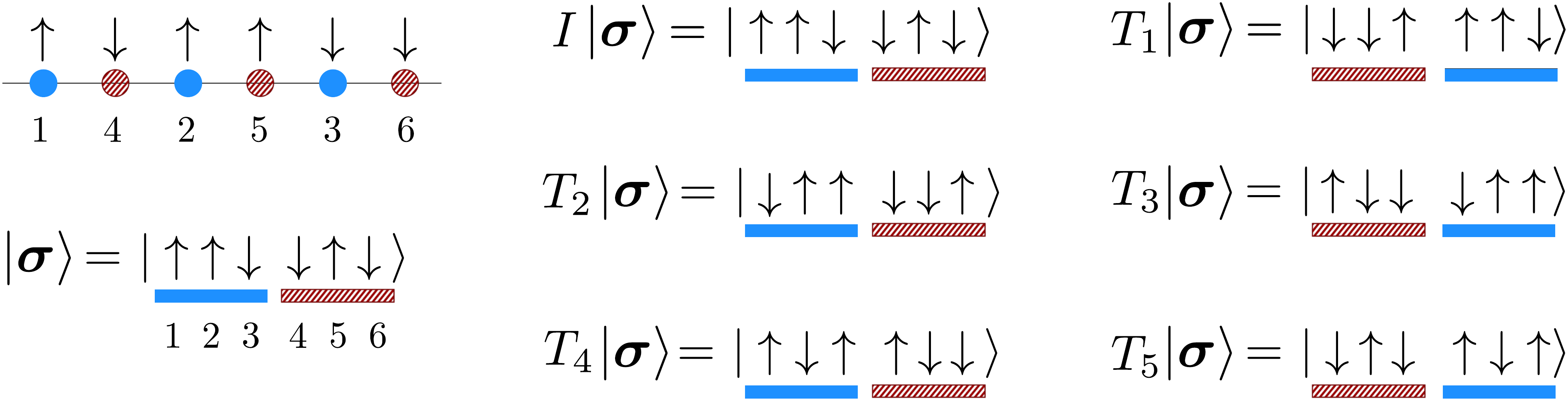}
  \caption{Two sublattice coding of the spin state $\ket{\bm{\sigma}}$
    on a six-site chain lattice and action of translational
    symmetries. The sites are enumerated such that site $1$-$3$ are on
    the blue (solid) sublattice $A$, $4$-$6$ on the red (dashed)
    sublattice $B$. The representative state with this enumeration of
    sites is given by
    $\ket{\tilde{\bm{\sigma}}} = T_1\ket{\bm{\sigma}} =
    \ket{\downarrow \downarrow \uparrow \uparrow \uparrow\downarrow}$.
    Notice, that the symmetries act on real space and thus the
    transformation of the basis states also depends on the numbering
    of sites.}
  \label{fig:sublatticetransformations}
\end{figure*}

Employing symmetries in ED computations amounts to block diagonalizing
the Hamiltonian. The blocks correspond to the irreducible
representations of the symmetry group and the procedure of block
diagonalization amounts to changing the basis of the Hilbert space to
\textit{symmetry-adapted basis states}. Here, we briefly review this
basis and recall some basic notions commonly used in this context. For
a more detailed introduction to this topic see
e.g. Refs.~\cite{Landau1988,Laeuchli2011}. In this manuscript, we only
consider one-dimensional representations of the symmetry
group. Consider a generic spin configuration on $N$ lattice sites with
local dimension $d$,
\begin{equation}
  \label{eq:spinconf_lsed}
  \ket{\bm{\sigma}} = \ket{\sigma_1,\ldots, \sigma_N}, \quad
  \sigma_i \in\{ 1, \ldots, d\}.
\end{equation}
The symmetry-adapted basis states $\ket{\bm{\sigma}_\rho}$ are given
by
\begin{align}
  \label{eq:symmetryadaptedwavefunction1d_lsed}
  \ket{\bm{\sigma}_\rho} \equiv \frac{1}{N_{\rho,\bm{\sigma}}}
  \sum\limits_{g \in \mathcal{G}} \chi_\rho(g)^* {g}\ket{\bm{\sigma}},
\end{align}
where $\mathcal{G}$ denotes a discrete symmetry group, $\rho$ a
one-dimensional representation of this group, $\chi_\rho(g)$ the
character of this representation evaluated at group element $g$, and
$N_{\rho,\bm{\sigma}}$ denotes the normalization constant of the state
$\ket{\bm{\sigma}_\rho}$. The set of basis state spin configurations
$\ket{\bm{\sigma}}$ is divided into \textit{orbits},
\begin{align}
  \label{eq:hilbertspaceorbits_lsed}
  \text{Orbit}(\ket{\bm{\sigma}}) = \{ {g} \ket{\bm{\sigma}}
  | g \in \mathcal{G}\}.
\end{align}
We define,
\begin{equation}
  \ket{\bm{\sigma}} < \ket{\bm{\sigma}^{\prime}} :\Leftrightarrow
  \integer(\ket{\bm{\sigma}}) < \integer(\ket{\bm{\sigma}^{\prime}}),
\end{equation}
where $\integer(\ket{\bm{\sigma}})$ denotes an integer value coding 
on the computer for the spin configuration $\ket{\bm{\sigma}}$.
The \textit{representative} $\ket{\tilde{\bm{\sigma}}}$ within each
orbit is given by the element with smallest integer value,
\begin{align}
  \label{eq:representativecondition_lsed}
  \ket{\tilde{\bm{\sigma}}} = g_{\bm{\sigma}}\ket{\bm{\sigma}}, \quad
  g_{\bm{\sigma}} = \argmin_{g \in \mathcal{G}}
  \,\text{int}( g\ket{\bm{\sigma}}).
\end{align}
The matrix element
$\braket{\tilde{\bm{\sigma}^{\prime}}_\rho | H_k | \tilde{\bm{\sigma}}_\rho}$ for
non-branching terms $ H_k $ for two symmetry-adapted basis states with
representation $\rho$ is given by
\begin{align}
  \label{eq:matrixelementrepresentatives_lsed}
  \braket{\tilde{\bm{\sigma}^{\prime}}_\rho | H_k | \tilde{\bm{\sigma}}_\rho} =
  \chi_\rho(g_{\bm{\sigma}^{\prime}})  \frac{N_{\rho,\bm{\sigma}^{\prime}}}{N_{\rho,\bm{\sigma}}}
  \braket{\bm{\sigma}^{\prime} | H_k | \tilde{\bm{\sigma}}}.
\end{align}

\section{Sublattice coding algorithm}
Evaluating the matrix elements
$\braket{\tilde{\bm{\sigma}^{\prime}}_\rho | H_k | \tilde{\bm{\sigma}}_\rho}$ in
\cref{eq:matrixelementrepresentatives_lsed} for all basis states
$\ket{\tilde{\bm{\sigma}}_\rho}$ and $\ket{\tilde{\bm{\sigma}^{\prime}}_\rho}$
efficiently is the gist of employing symmetries in ED computations.
In an actual implementation on the computer we need to perform the
following steps:
\begin{itemize}
\item Apply the non-branching term $H_k$ on the representative state
  $\ket{\tilde{\bm{\sigma}}}$. This yields a possibly
  non-representative state $\ket{\bm{\sigma}^{\prime}}$. From this, we can
  compute the factor $\braket{\bm{\sigma}^{\prime} | H_k | \tilde{\bm{\sigma}}}$.
\item Find the representative $\ket{\tilde{\bm{\sigma}^{\prime}}}$ of
  $\ket{\bm{\sigma}^{\prime}}$ and determine the group element $g_{\bm{\sigma}^{\prime}}$
  such that
  $\ket{\tilde{\bm{\sigma}^{\prime}}} = {g}_{\bm{\sigma}^{\prime}}\ket{\bm{\sigma}^{\prime}}$. This
  yields the factor $\chi_\rho(g_{\bm{\sigma}^{\prime}})$.
\item Know the normalization constants $N_{\rho,\bm{\sigma}^{\prime}}$ and
  $N_{\rho,\bm{\sigma}}$. These are usually computed when creating a
  list of all representatives and stored in a separate list.
\end{itemize}
The problem of finding the representative $\ket{\tilde{\bm{\sigma}}}$
of a given state $\ket{\bm{\sigma}}$ and its corresponding symmetry
${g}_{\bm{\sigma}}$ turns out to be the computational bottleneck of ED
in a symmetrized basis. It is thus desirable to solve this problem
fast and memory efficient. There are two straightforward approaches to
solving this problem:
\begin{itemize}
\item Apply all symmetries directly to $\ket{\bm{\sigma}^{\prime}}$ to find the
  minimizing group element $g_{\bm{\sigma}^{\prime}}$,
  \begin{equation}
    g_{\bm{\sigma}^{\prime}} = \argmin_{g \in \mathcal{G}} \,
    \text{int}(g\ket{\bm{\sigma}^{\prime}}).
  \end{equation}
  This method does not have any memory overhead but is computationally
  slow since all symmetries have to be applied to the given state
  $\ket{\bm{\sigma}^{\prime}}$.
\item For every state $\ket{\bm{\sigma}}$ we store
  $\ket{\tilde{\bm{\sigma}}}$ and $g_{\bm{\sigma}}$ in a lookup
  table. While this is very fast computationally, the lookup table for
  storing all representatives grows exponentially in the system size.
\end{itemize}
The key to solving the representative search problem adequately is to
have an algorithm that is almost as fast as a lookup table, where
memory requirements are within reasonable bounds. This problem has
already been addressed by several authors \cite{Lin1990,
  Weisse2013}. The central idea in these so-called \textit{sublattice
  coding techniques} is to have a lookup table for the representatives
on a sublattice of the original lattice and combine the information of
the sublattice representatives to compute the total
representative. These ideas were first introduced in \cite{Lin1990,
  Weisse2013, Schulz1996}.  In the following paragraphs, we explain
the basic idea behind these algorithms and propose a flexible
extension to arbitrary geometries and number of sublattices.

\begin{figure*}[t!]
  \centering
  \begin{minipage}[c]{0.3\textwidth}
    \centering \includegraphics[width=0.5\textwidth]{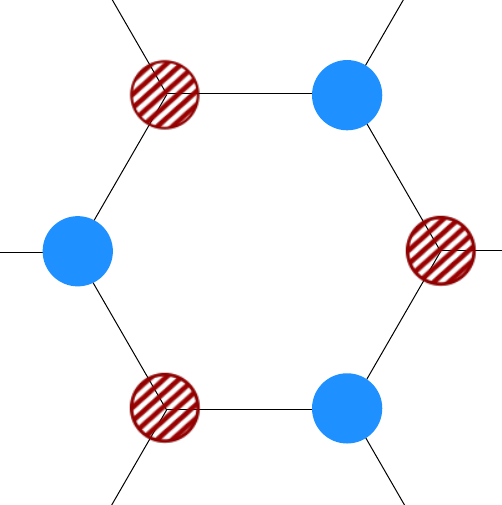}
  \end{minipage}%
  \quad
  \begin{minipage}[c]{0.3\textwidth}
    \includegraphics[width=0.8\textwidth]{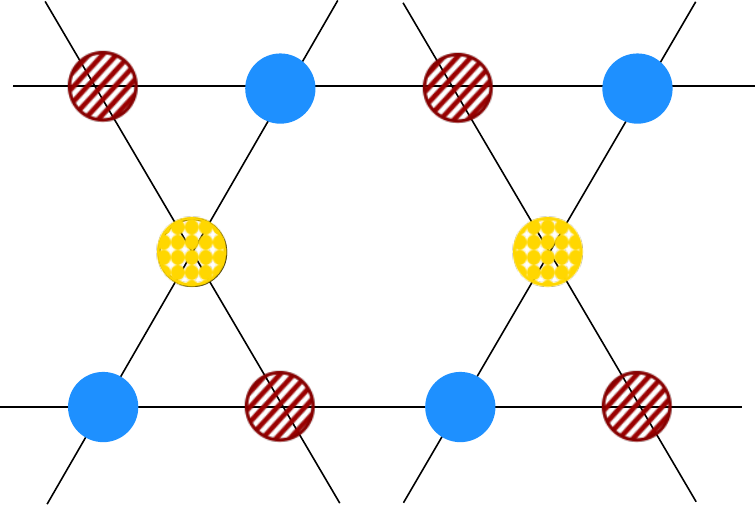}
  \end{minipage}
  \quad
  \begin{minipage}[c]{0.3\textwidth}
    \quad \includegraphics[width=0.7\textwidth]{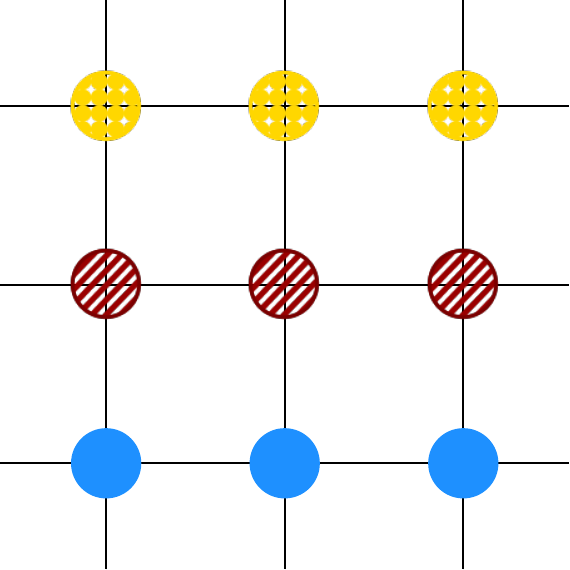}
  \end{minipage}

  \caption{Sublattice orderings for several common lattices. The
    sublattices are distinguished by different colors (shadings).
    \textit{Left}: Two sublattice ordering in a honeycomb lattice. The
    sublattices are stable with respect to all spatial
    symmetries. \textit{Middle}: Three sublattice ordering on a kagome
    lattice. The sublattices are stable with respect to all spatial
    symmetries. \textit{Right}: Three sublattice ordering on a square
    lattice. The sublattices are stable with respect to all
    translational symmetries, horizontal and vertical reflections,
    $180^\circ$ rotations but not with respect to $90^\circ$ rotations
    or diagonal reflections.}
  \label{fig:sublatticeorderings}
\end{figure*}
\subsection{Sublattice coding on two sublattices}
For demonstration purposes, we consider a simple translationally
invariant spin-$1/2$ system on a six-site chain lattice with periodic
boundary conditions.  The lattice is divided into two sublattices as
in \cref{fig:sublatticetransformations}. The even sites form the
sublattice $A$ and the odd sites form the sublattice $B$. We enumerate
the sites such that the sites $1$ to $3$ are in sublattice $A$ and the
sites $4$ to $6$ are in sublattice $B$. We choose the integer
representation of a state $\ket{\bm{\sigma}}$ such that the most
significant bits are formed by the spins in sublattice $A$.  The
symmetry group we consider consists of the six translations on the
chain
\begin{equation}
  \mathcal{G} = \{ \text{Id}, T, T_2, T_3, T_4, T_5 \},
\end{equation}
where $T_n$ denotes the translation by $n$ lattice sites.  The
splitting of the lattice into two sublattices is stable in the sense
that every symmetry element $g \in \mathcal{G} $ either maps the $A$
sublattice to $A$ and the $B$ sublattice to $B$ or the $A$ sublattice
to $B$ and the $B$ sublattice to $A$.  We call this property
\textit{sublattice stability}. It is both a property of the partition
of our lattice into sublattices and the symmetry group. Hence, the
symmetry group is composed of two kinds of symmetries
\begin{align}
  \begin{split}
    \mathcal{G}_A &\equiv \{ g \in \mathcal{G} \, ; \quad g \text{
      maps sublattice } A \text{ onto } A \},
    \\
    \mathcal{G}_B &\equiv \{ g \in \mathcal{G} \, ; \quad g \text{
      maps sublattice } B \text{ onto } A \}.
  \end{split}
\end{align}
We denote by $\ket{\bm{\sigma}}_A$ (resp. $\ket{\bm{\sigma}}_B$) the
state restricted to sublattice $A$ (resp. $B$) and define the
\textit{sublattice representatives},
\begin{align}
  \label{eq:sublatreps2sl}
  \begin{split}
    \texttt{Rep}_A(\ket{\bm{\sigma}}_A) &\equiv
    h_A\ket{\bm{\sigma}}_A, \quad h_A = \argmin_{g \in \mathcal{G}_A}
    \, \text{int}(g \ket{\bm{\sigma}}_A),
    \\
    \texttt{Rep}_B(\ket{\bm{\sigma}}_B) &\equiv
    h_B\ket{\bm{\sigma}}_B, \quad h_B = \argmin_{g \in \mathcal{G}_B}
    \, \text{int}(g \ket{\bm{\sigma}}_B),
  \end{split}
\end{align}
and the \textit{representative symmetries},
\begin{align}
  \label{eq:repsyms2sl}
  \begin{split}
    \texttt{Sym}_A(\ket{\bm{\sigma}}_A) &\equiv \{ g \in \mathcal{G}_A
    \, ; \quad {g}\ket{\bm{\sigma}}_A =
    \texttt{Rep}_A(\ket{\bm{\sigma}}_A) \},
    \\
    \texttt{Sym}_B(\ket{\bm{\sigma}}_B) &\equiv \{ g \in \mathcal{G}_B
    \, ; \quad {g}\ket{\bm{\sigma}}_B =
    \texttt{Rep}_B(\ket{\bm{\sigma}}_B) \}.
  \end{split}
\end{align}
\bigbreak\noindent Let again
$\ket{\tilde{\bm{\sigma}}} = g_{\bm{\sigma}}\ket{\bm{\sigma}}$, where
$\ket{\tilde{\bm{\sigma}}}$ is the representative of
$\ket{\bm{\sigma}}$.  The minimizing symmetry $g_{\bm{\sigma}}$ can
only be an element of $\texttt{Sym}_A(\ket{\bm{\sigma}}_A)$ if
$\texttt{Rep}_A(\ket{\bm{\sigma}}_A) \leq
\texttt{Rep}_B(\ket{\bm{\sigma}}_B)$, or vice versa. Put differently,
\begin{equation}
  \label{eq:sublatticerepslogic}
  \texttt{Rep}_B(\ket{\bm{\sigma}}_B) < \texttt{Rep}_A(\ket{\bm{\sigma}}_A)
  \quad\Rightarrow \quad
  g_{\bm{\sigma}} \notin \texttt{Sym}_A(\ket{\bm{\sigma}}_A).  
\end{equation}
Otherwise, any symmetry element in
$\texttt{Rep}_B(\ket{\bm{\sigma}}_B)$ would yield a smaller integer
value than $g_{\bm{\sigma}}$. This is the core idea behind the
sublattice coding technique. We store
$\texttt{Rep}_{A,B}(\ket{\bm{\sigma}}_{A,B})$ for every substate
$\ket{\bm{\sigma}}_{A,B}$ in a lookup table together with
$\texttt{Sym}_{A,B}(\ket{\bm{\sigma}}_{A,B})$. In a first step, we
determine the sublattice representative with smallest most significant
bits. Then we apply the representative symmetries to
$\ket{\bm{\sigma}}$ in order to determine the true representative
$\ket{\tilde{\bm{\sigma}}}$. The number of representative symmetries
$|\texttt{Sym}_{A,B}(\ket{\bm{\sigma}}_{A,B})|$ is typically much
smaller than the total number of symmetries $|\mathcal{G}|$. The
following example illustrates the idea and shows how to compute the
representative given the information about sublattice representatives
and representative symmetries.

\begin{figure*}[t!]
  \centering \includegraphics[width=0.8\textwidth]{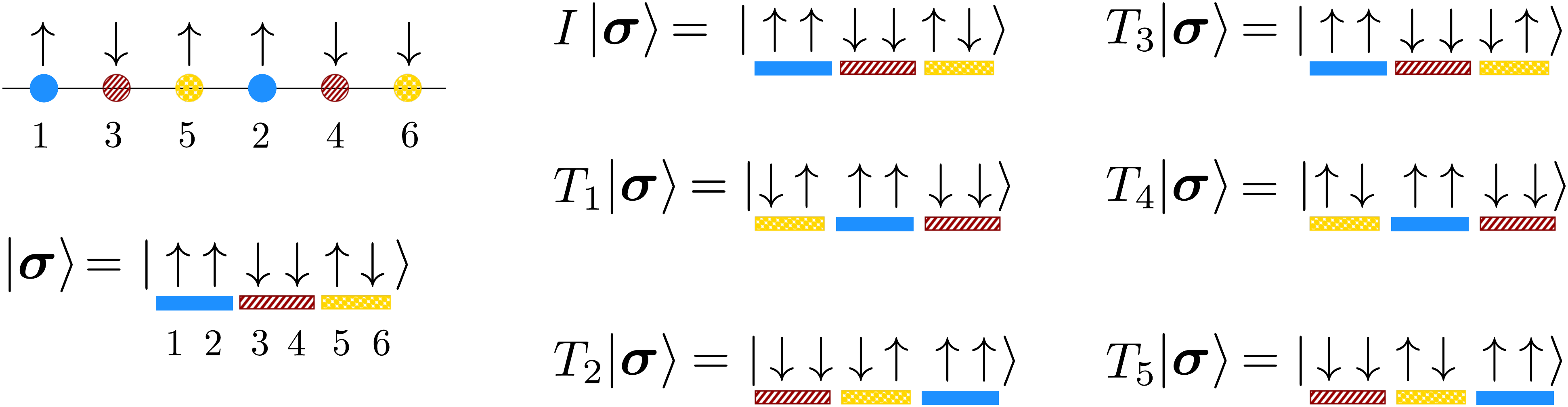}
  \caption{Three sublattice coding of the spin state
    $\ket{\bm{\sigma}}$ on a six-site chain lattice and action of
    translation symmetries. The sites are enumerated such that site
    $1$ and $2$ are on sublattice $A$ (blue, solid), $3$,$4$ on $B$
    (red, dashed) and $5$,$6$ on $C$ (yellow, dotted). The
    representative state with this enumeration of sites is given by
    $\ket{\tilde{\bm{\sigma}}} = T_2\ket{\bm{\sigma}} =
    \ket{\downarrow \downarrow \downarrow \uparrow \uparrow
      \uparrow}$}
  \label{fig:threesublatticetransformations}
\end{figure*}

\paragraph*{Example} We consider the state
$\ket{\bm{\sigma}} = \ket{\uparrow \uparrow \downarrow \downarrow
  \uparrow \downarrow}$ on a six-site chain lattice as in
\cref{fig:sublatticetransformations}. Notice that the sites are not
enumerated from left to right but such that sites $1$ to $3$ belong to
the sublattice $A$ and sites $4$ to $6$ belong to sublattice $B$. The
states restricted on the sublattices are
$\ket{\bm{\sigma}}_A = \ket{\uparrow \uparrow \downarrow}$ and
$\ket{\bm{\sigma}}_B = \ket{\downarrow \uparrow \downarrow}$. The
action of the sublattice symmetries
\begin{align}
  \begin{split}
    \mathcal{G}_A &\equiv \{ \text{Id}, T_2, T_4 \}, \\
    \mathcal{G}_B &\equiv \{ T_1, T_3, T_5 \}, \\
  \end{split}
\end{align}
on $\ket{\bm{\sigma}}$ is shown in
\cref{fig:sublatticetransformations}. From this, we compute
the sublattice representatives as in \cref{eq:sublatreps2sl},
\begin{align}
  \begin{split}
    \texttt{Rep}_A(\ket{\bm{\sigma}}_A) &=
    \ket{\downarrow\uparrow\uparrow},
    \\
    \texttt{Rep}_B(\ket{\bm{\sigma}}_B) &=
    \ket{\downarrow\downarrow\uparrow},
  \end{split}
\end{align}
whose integer values are given by
\begin{align}
  \begin{split}
    \integer(\texttt{Rep}_A(\ket{\bm{\sigma}}_A)) &= (011)_2 = 3, \\
    \integer(\texttt{Rep}_B(\ket{\bm{\sigma}}_B)) &= (001)_2 = 1.
  \end{split}
\end{align}
Since
$\texttt{Rep}_B(\ket{\bm{\sigma}}_B) <
\texttt{Rep}_A(\ket{\bm{\sigma}}_A)$ the symmetry $g_{\bm{\sigma}}$
yielding the total representative $\ket{\tilde{\bm{\sigma}}}$ must be
contained in
\begin{equation}
  \texttt{Sym}_B(\ket{\bm{\sigma}}_B) = \{ T_1 \},      
\end{equation}
which in this case just contains a single element, namely $T_1$.
Consequently, the representative $ \ket{\tilde{\bm{\sigma}}} $ is
given by
\begin{equation}
  \ket{\tilde{\bm{\sigma}}} =  T_1 \ket{\bm{\sigma}} =
  \ket{\downarrow \downarrow \uparrow  \uparrow \uparrow \downarrow}.      
\end{equation}
\bigbreak
\paragraph*{Lookup tables} If the quantities
$\texttt{Rep}_{A,B}(\ket{\bm{\sigma}}_{A,B})$ and
$\texttt{Sym}_{A,B}(\ket{\bm{\sigma}}_{A,B})$ are now stored in a
lookup table, this computation can be done very efficiently. Notice
that instead of having to store $2^N$ entries in the lookup table for
the representative we only need four lookup tables of order
$\mathcal{O}(2^{N/2})$, two for the quantities 
$\texttt{Rep}_{A}(\ket{\bm{\sigma}}_{A})$ and $\texttt{Rep}_{B}(\ket{\bm{\sigma}}_{B})$, 
and two for $\texttt{Sym}_{A}(\ket{\bm{\sigma}}_{A})$ and
$\texttt{Sym}_{B}(\ket{\bm{\sigma}}_{B})$. On larger system sizes the 
difference between memory requirements of order $\mathcal{O}(2^{N})$ and
$\mathcal{O}(2^{N/2})$ is substantial.

To further speed up computations we also create lookup tables to store
the action of each symmetry $g \in \mathcal{G}$ on a substate
$\ket{\bm{\sigma}}_A$,
\begin{align}
  \begin{split}
    \texttt{SymmetryAction}_A(g, \ket{\bm{\sigma}}_A) &=
    {g}\ket{\bm{\sigma}}_A,
    \\
    \texttt{SymmetryAction}_B(g, \ket{\bm{\sigma}}_B) &=
    {g}\ket{\bm{\sigma}}_B.
  \end{split}
\end{align}
With this information, we can efficiently apply symmetries to a given
spin configuration by looking up the action of $g$ on the respective
substate and combining the results. The memory requirement for these
lookup tables is $\mathcal{O}(N_{\text{sym}}2^{N/2})$, where
$N_{\text{sym}} = |\mathcal{G}|$. This can be reduced by generalizing
the sublattice coding algorithm to multiple sublattices, as explained in the
following section. In that case, 
$2N_{\text{sublat}}$ lookup tables of size $\mathcal{O}(2^{N/N_{\text{sublat}}})$
are required for storing the
sublattice representatives $\texttt{Rep}_X(\ket{\bm{\sigma}}_X)$
and representative symmetries $\texttt{Sym}_X(\ket{\bm{\sigma}}_X)$, as 
defined in \cref{eq:sublatrepdef,eq:repsymdef}, respectively.
$N_{\text{sublat}}$ denotes the number of sublattices. For storing the 
action of each symmetry $\texttt{SymmetryAction}_X(g, \ket{\bm{\sigma}}_X)$
as in \cref{eq:symactiondef} we further need $N_{\text{sublat}}$ lookup tables
of size $\mathcal{O}(N_{\text{sym}}2^{N/N_{\text{sublat}}})$.

\subsection{Generic sublattice coding algorithm}

We start by discussing how we subdivide a lattice $\Lambda$ into
$N_{\text{sublat}}$ sublattices. The basic requirement is that every
symmetry group element either only operates within the sublattices or
exchanges sublattices. We do not allow for symmetry elements that
split up a sublattice onto different sublattices. Therefore we make
the following definition:

\paragraph*{Definition (Sublattice stability)}
A decomposition,
\begin{equation}
  \Lambda = \bigcupdot\limits_{X=1}^{N_{\text{sublat}}} \Lambda_X ,
\end{equation}
of a lattice $\Lambda$ with symmetry group $\mathcal{G}$ into
$N_{\text{sublat}}$ disjoint sublattices $\Lambda_X$ is called
\textit{sublattice stable} if every $g\in \mathcal{G}$ maps each
$\Lambda_X$ onto exactly one (possibly different) $\Lambda_Y$,
i.e. for all $g\in \mathcal{G}$ and all $\Lambda_X$ there exists a
$\Lambda_Y$ such that
\[
  g(\Lambda_X) = \Lambda_Y.
\]
The set $\Lambda_X$ is called the $X$-\textit{sublattice} of
$\Lambda$.

\bigbreak The notion of \textit{sublattice stability} is illustrated
in \cref{fig:sublatticeorderings}. The sublattices $\Lambda_X$ are
drawn in different colors (shadings). A translation by one unit cell
keeps the sublattices of the honeycomb lattice invariant whereas a
$60^\circ$ rotation exchanges the sublattices.  For the kagome lattice
in \cref{fig:sublatticeorderings} a $60^\circ$ rotation around a
hexagon center for example cyclically permutes the three
sublattices. One checks that for both the honeycomb and the kagome
lattice in \cref{fig:sublatticeorderings} all translational as well as
all point group symmetries are sublattice stable, so different color
(shading) sublattices are mapped onto each other. This is different
for the square lattice in \cref{fig:sublatticeorderings}. Still here
all translational symmetries just permute the sublattices, but a
$90^\circ$ rotation splits up a sublattice into different
sublattices. Nevertheless, a $180^\circ$ rotation keeps the
sublattices stable, similarly a vertical or horizontal
reflection. Therefore, only the reduced point group $\mathrm{D}2$
instead of the full $\mathrm{D}4$ point group for the square lattice
fulfills the sublattice stability condition in this
case. $\mathrm{D}2$ and $\mathrm{D}4$ denote the dihedral groups of
order $4$ and $8$ with two- and four-fold rotations and
reflections. Note, that for a square lattice a two or four sublattice
decomposition for which the full $\mathrm{D}4$ point group is
sublattice stable can be chosen instead. The choice of this particular
sublattice decomposition just serves illustrational purposes.

From the definition of sublattice stability, it is clear that the
total number of sites $N$ has to be divisible by the number of
sublattices $N_{\text{sublat}}$.  The numbering of the lattice sites
is chosen such that the lattice sites from
$(X-1)N/N_{\text{sublat}} + 1$ to $XN/N_{\text{sublat}}$ belong to
sublattice $X$. We choose the most significant bits in the integer
representation to be the bits on sublattice $1$. Similar as in the
previous section we define the following quantities
\paragraph*{Definition} \label{def:lookuptables} For every sublattice
$\Lambda_X$ we define the following notions:
\begin{itemize}
\item \textit{sublattice symmetries}:
  \begin{equation} \label{eq:sublatsymdef} \mathcal{G}_X \equiv \{ g
    \in \mathcal{G} \, | \, g \text{ maps sublattice } X \text{ onto
      sublattice } 1 \}.
  \end{equation}
\item \textit{sublattice representative}:
  \begin{equation} \label{eq:sublatrepdef}
    \texttt{Rep}_X(\ket{\bm{\sigma}}_X) \equiv h_X\ket{\bm{\sigma}}_X,
    \quad h_X = \argmin_{g \in \mathcal{G}_X} \, \text{int}(g
    \ket{\bm{\sigma}}_X),
  \end{equation}
  where $\ket{\bm{\sigma}}_X$ denotes the substate of
  $\ket{\bm{\sigma}}$ restricted on sublattice $\Lambda_X$.
\item \textit{representative symmetries}:
  \begin{equation} \label{eq:repsymdef}
    \texttt{Sym}_X(\ket{\bm{\sigma}}_X) \equiv \{ g \in \mathcal{G}_X
    \, | \, {g}\ket{\bm{\sigma}}_X =
    \texttt{Rep}_X(\ket{\bm{\sigma}}_X) \}.
  \end{equation}
\item \textit{sublattice symmetry action}:
  \begin{equation} \label{eq:symactiondef}
    \texttt{SymmetryAction}_X(g, \ket{\bm{\sigma}}_X) =
    {g}\ket{\bm{\sigma}}_X.
  \end{equation}
\end{itemize}
The symmetries in $\mathcal{G}_X$ map the sublattice $X$ onto the most
significant bits. Therefore, the symmetry that minimizes the integer
value in the orbit must be contained in the representative symmetries
of a minimal sublattice representative, i.e.
\begin{equation}
  g_{\bm{\sigma}} = \argmin\limits_{g \in \mathcal{G}} {g}\ket{\bm{\sigma}}
  \Rightarrow
  g_{\bm{\sigma}} \in \bigcup\limits_{\substack{Y\text{, \texttt{Rep}}_Y(\ket{\bm{\sigma}}_Y)
      \\ \text{minimal}}}{\texttt{Sym}}_Y(\ket{\bm{\sigma}}_Y). 
\end{equation}
To find the minimizing symmetry $g_{\bm{\sigma}}$, we only have to
check the symmetries yielding the minimal sublattice
representative. The quantities $\texttt{Rep}_X(\ket{\bm{\sigma}}_X)$
and $\texttt{Sym}_X(\ket{\bm{\sigma}}_X)$ are stored in lookup tables,
whose size scales as $\mathcal{O}(2^{N/N_{\text{sublat}}})$. In order
to quickly apply the symmetries, we can additionally store
$\texttt{SymmetryAction}_X(g, \ket{\bm{\sigma}}_X)$ in another lookup
table. The memory cost of doing so scales as
$\mathcal{O}(N_{\text{sym}}2^{N/N_{\text{sublat}}})$ and thus requires
the most memory. The generic sublattice coding algorithm consists of
two parts. The preparation of the lookup tables is shown as pseudocode
in \cref{alg:sublatticecodingprep}. The pseudocode of the actual
algorithm for finding the representative using the lookup tables is
shown in \cref{alg:sublatticecodingfindrep}.

% \begin{algorithm}[H]
%   \textbf{Preparation: }\\
%   \For{ sublattice $X$} \For{ substate $\ket{\bm{\sigma}}_X$}
%   compute the sublattice representative \cref{eq:sublatrepdef}, store it to $\texttt{Rep}_X(\ket{\bm{\sigma}}_X)$ \\
%   compute the representative symmetries \cref{eq:repsymdef}, store them to $\texttt{Sym}_X(\ket{\bm{\sigma}}_X)$ \\
%   \EndFor \EndFor \For{ sublattice $X$}{ \For{ substate
%   $\ket{\bm{\sigma}}_X$}{ \For{ symmetry $g \in \mathcal{G}$}{
%   compute ${g}\ket{\bm{\sigma}}_X$ and store it to
%   $\texttt{SymmetryAction}_X(g, \ket{\bm{\sigma}}_X)$ } \EndFor
%   \EndFor \EndFor

%   \caption{Preparation step for initializing the lookup tables. The
%   size of $\texttt{Rep}_X(\ket{\bm{\sigma}}_X)$ and
%   $\texttt{Sym}_X(\ket{\bm{\sigma}}_X)$ scales as
%   $\mathcal{O}(2^{N/N_{\text{sublat}}})$ whereas the size of the
%   lookup table $\texttt{SymmetryAction}_X(g, \ket{\bm{\sigma}}_X)$
%   scales as $\mathcal{O}(N_{\text{sym}}2^{N/N_{\text{sublat}}})$.}
% \end{algorithm}
\begin{figure}
  \begin{algorithm}[H]
    \caption{Preparation of lookup tables for sublattice coding
      algorithm}
    \begin{algorithmic}[0]
      \ForEach{substate $\ket{\bm{\sigma}_X}$} \ForEach{sublattice
        $X$} \State compute $\texttt{Rep}_X(\ket{\bm{\sigma}}_X)$
      \cref{eq:sublatrepdef}, store it \State compute
      $\texttt{Sym}_X(\ket{\bm{\sigma}}_X)$ \cref{eq:repsymdef}, store
      them \ForEach{ symmetry $g \in \mathcal{G}$} \State compute
      $\texttt{SymmetryAction}_X(g, \ket{\bm{\sigma}}_X)$, store it
      \EndForEach \EndForEach \EndForEach
    \end{algorithmic}
    \label{alg:sublatticecodingprep}
  \end{algorithm}
\end{figure}

\begin{figure}
  \begin{algorithm}[H]
    \caption{Sublattice coding algorithm for finding the
      representative.}
    \begin{algorithmic}[0]
      \Require state $\ket{\bm{\sigma}}$ \Ensure representative
      $\ket{\tilde{\bm{\sigma}}}$ and $g_{\bm{\sigma}}$ \State
      Determine
      $\texttt{MinRep} = \min\limits_{X} \left\{
        \texttt{Rep}_X(\ket{\bm{\sigma}}_X)\right\}$ \State Set
      $\ket{\tilde{\bm{\sigma}}} = +\infty$ \ForEach{ sublattice $Y$
        with $\texttt{Rep}_Y(\ket{\bm{\sigma}}_Y) = \texttt{MinRep}$}
      \ForEach{ symmetry $ g \in \texttt{Sym}_Y(\ket{\bm{\sigma}}_Y)$}
      \State compute ${g} \ket{\bm{\sigma}}$ from
      $\texttt{SymmetryAction}_X(g, \ket{\bm{\sigma}}_X)$
      \If{${g} \ket{\bm{\sigma}} < \ket{\tilde{\bm{\sigma}}}$} \State
      $\ket{\tilde{\bm{\sigma}}} \leftarrow {g} \ket{\bm{\sigma}} $
      \State $g_{\bm{\sigma}} \leftarrow {g}$ \EndIf \EndForEach
      \EndForEach \State \Return $\ket{\tilde{\bm{\sigma}}}$,
      $g_{\bm{\sigma}}$
    \end{algorithmic}
    \label{alg:sublatticecodingfindrep}
  \end{algorithm}
\end{figure}

\paragraph*{Example} We consider the same state on a six-site chain
lattice as in \cref{fig:sublatticetransformations}, but now using a
three sublattice decomposition in
\cref{fig:threesublatticetransformations}. We call the blue (solid)
sublattice the $A$ sublattice, the red (dashed) $B$ and the yellow
(dotted) $C$. Notice, that due to different sublattice structure the
labeling of the real space sites is different from the two sublattice
case. In the three sublattice case, we are now given the state
\begin{equation}
  \ket{\bm{\sigma}} = \ket{ \uparrow \uparrow \downarrow \downarrow \uparrow \downarrow}.
\end{equation}
Its substates are
\begin{align}
  \begin{split}
    \ket{\bm{\sigma}}_A = \ket{ \uparrow \uparrow}, \\
    \ket{\bm{\sigma}}_B = \ket{ \downarrow \downarrow}, \\
    \ket{\bm{\sigma}}_C = \ket{ \uparrow \downarrow},
  \end{split}
\end{align}
with corresponding sublattice representatives
\begin{align}
  \begin{split}
    \texttt{Rep}_A(\ket{\bm{\sigma}}_A) = \ket{ \uparrow \uparrow}, \\
    \texttt{Rep}_B(\ket{\bm{\sigma}}_B) = \ket{ \downarrow \downarrow}, \\
    \texttt{Rep}_C(\ket{\bm{\sigma}}_C) = \ket{ \downarrow \uparrow},
  \end{split}
\end{align}
and representative symmetries
\begin{align}
  \begin{split}
    \texttt{Sym}_A(\ket{\bm{\sigma}}_A) = \{ I, T_3\}, \\
    \texttt{Sym}_B(\ket{\bm{\sigma}}_B) = \{ T_2, T_5\}, \\
    \texttt{Sym}_C(\ket{\bm{\sigma}}_C) = \{ T_1\}.
  \end{split}
\end{align}
The minimal sublattice representative $\texttt{MinRep}$ as in
\cref{alg:sublatticecodingfindrep} is given by
\begin{equation}
  \texttt{MinRep} = \texttt{Rep}_B(\ket{\bm{\sigma}}_B) =
  \ket{ \downarrow \downarrow}.
\end{equation}
The minimizing symmetry must now be in
$\texttt{Sym}_B(\ket{\bm{\sigma}}_B) = \{ T_2, T_5\}$.  We see that
\begin{equation}
  T_2\ket{\bm{\sigma}} =
  \ket{\downarrow \downarrow \downarrow \uparrow \uparrow \uparrow} <
  T_5\ket{\bm{\sigma}} =
  \ket{\downarrow \downarrow \uparrow \downarrow \uparrow \uparrow}. 
\end{equation}
Therefore, the representative $\ket{\tilde{\bm{\sigma}}}$ is given by
\begin{equation}
  \ket{\tilde{\bm{\sigma}}} = \ket{\downarrow \downarrow \downarrow \uparrow \uparrow \uparrow},
\end{equation}
with the minimizing symmetry $g_{\bm{\sigma}} = T_2$.  Notice, that
this state differs from the one found in the two sublattice example
since the labeling of the sites changes the integer representation of
a state and thus the definition of the representative. Once a given
labeling of sites is fixed the representative is of course unique.

\section{Distributed and hybrid memory parallelization}

For reaching larger system sizes in ED computations a proper balance
between memory requirements and computational costs has to be
found. There are two major approaches when applying the Lanczos
algorithm. The Hamiltonian matrix can either be stored in memory in
some sparse-matrix format or generated on-the-fly every time a
matrix-vector multiplication is performed. Storing the matrix is
usually faster, yet memory requirements are higher.  This approach is
for example pursued by the software package \texttt{SPINPACK}
\cite{Spinpack}.  A matrix-free implementation of the Lanczos
algorithm usually needs more computational time since the matrix
generation, especially in a symmetrized basis can be expensive. Of
course, the memory cost is drastically reduced since only a few
vectors of the size of the Hilbert space have to be stored. It turns
out that on current supercomputing infrastructures the main limitation
in going to larger system sizes is indeed the memory requirements of
the computation. It is thus often favorable to use a slower
matrix-free implementation, as done by the software package
${\mathcal H}\Phi$ \cite{Kawamura2017}, for example. Due to this
reasons, we also choose the matrix-free approach.

The most computational time in the Lanczos algorithm is used in the
matrix-vector multiplication. The remaining types of operations are
scalar multiplications, dot products of Lanczos vectors or the
diagonalization of the $T$-matrix which are usually of negligible
computational cost. Today's largest supercomputers are typically
distributed memory machines, where every process only has direct
access to a small part of the total memory. It is thus a nontrivial
task to distribute data onto several processes and implement
communication amongst them once remote memory has to be
accessed. Also, when scaling the software to a larger amount of
processes load balancing becomes important. The computational work
should be evenly distributed amongst the individual processes in order
to avoid waiting times in communication. In the following, we explain
how we achieve this goal in our implementation using the Message
Passing Protocol (MPI).

\paragraph*{Matrix-vector multiplication}

\begin{figure}[t!]
  \centering \includegraphics[width=\columnwidth]{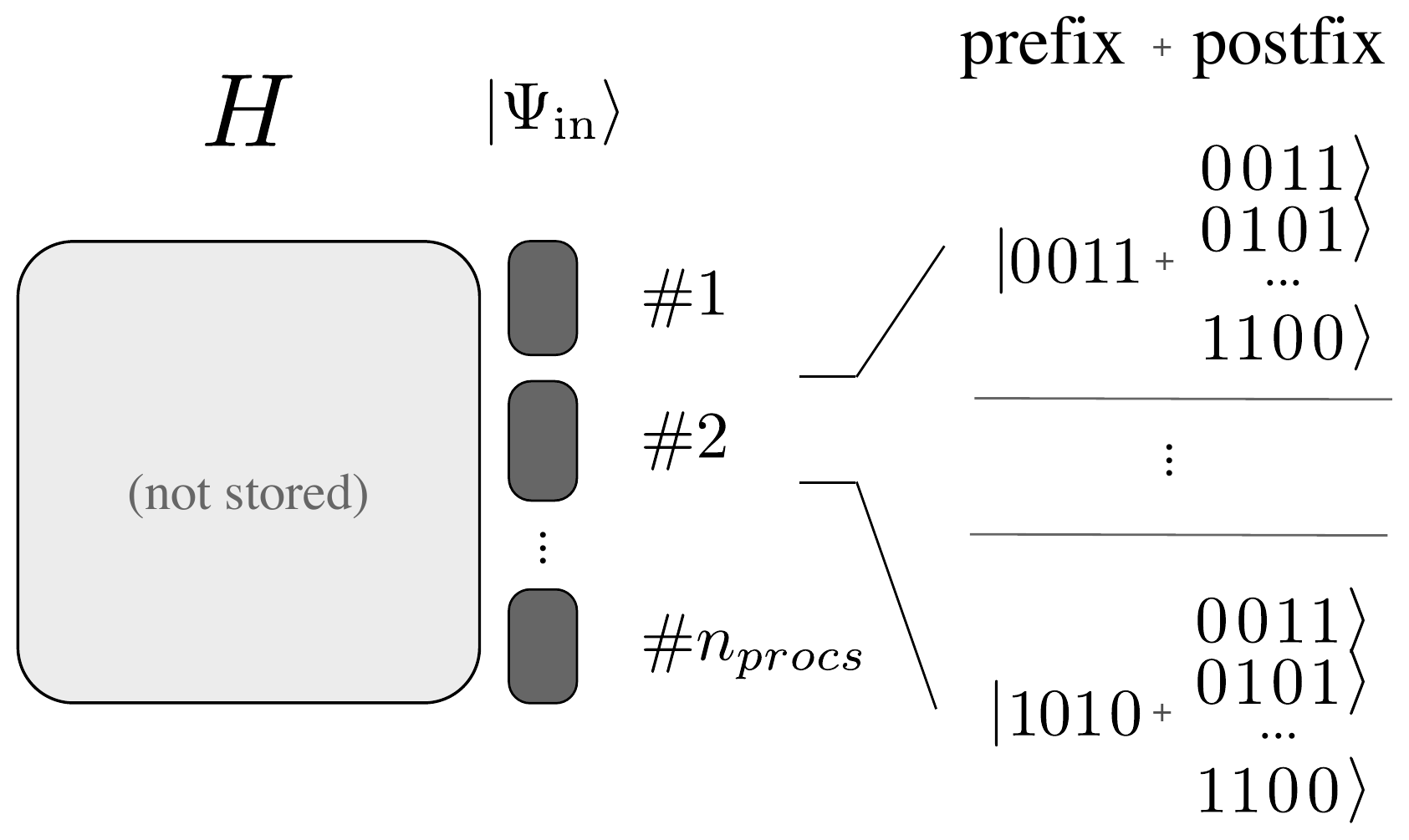}
  \caption{Storage layout of the distributed Hilbert space. The
    prefixes are randomly distributed amongst the MPI processes using
    a hash function. States with same prefixes are mapped to the same
    process. Within a process, the states are ordered
    lexicographically. The Hamiltonian matrix is not stored.}
  \label{fig:edparallelization}
\end{figure}

The Hamiltonian can be written a sum of non-branching terms,
\begin{align}
  H = \sum \limits_k H_k.
\end{align}
To perform the full matrix-vector multiplication we compute the
matrix-vector multiplication for the non-branching terms $H_k$ and add
up the results,
\begin{equation}
  \label{eq:branchingsum}
  H\ket{\psi} = \sum  \limits_k H_k \ket{\psi}.
\end{equation}
We denote by
\begin{equation}
  \label{eq:sigmabasis}
  \{ \ket{\bm{\sigma}_i} \}, \quad i=1,\ldots,D \quad ,
\end{equation}
a (possibly symmetry-adapted) basis of the Hilbert space.  A wave
function $\ket{\psi}$ is represented on the computer by storing its
coefficients $\braket{\bm{\sigma}_i| \psi}$.  Given an input vector,
\begin{equation}
  \label{eq:inputvector}
  \ket{\psi_{\text{in}}} = \sum\limits_{i=1}^D
  \braket{\bm{\sigma}_i| \psi_{\text{in}}} \ket{\bm{\sigma}_i},
\end{equation}
we want to compute the coefficients
$\braket{\bm{\sigma}_i| \psi_{\text{out}}}$ in
\begin{equation}
  H_k\ket{\psi_{\text{in}}} = \ket{\psi_{\text{out}}}.
\end{equation}
The resulting output vector $\ket{\psi_{\text{out}}}$ is given by
\begin{align}
  \begin{split}
    \ket{\psi_{\text{out}}} &= \sum\limits_{i=1}^D
    \braket{\bm{\sigma}_i| \psi_{\text{out}}} \ket{\bm{\sigma}_i}=
    \sum\limits_{i=1}^D
    \braket{\bm{\sigma}_i| H_k|\psi_{\text{in}}} \ket{\bm{\sigma}_i} \\
    & = \sum\limits_{i,j=1}^D c_k(\bm{\sigma}_j)\braket{\bm{\sigma}_j|
      \psi_{\text{in}}} \braket{\bm{\sigma}_i|\bm{\sigma}_j^\prime}
    \ket{\bm{\sigma}_i},
  \end{split}
      \label{eq:outputvector}  
\end{align}
where $ c_k(\bm{\sigma}_j)$ and $\ket{\bm{\sigma}_j^\prime}$ are given
by
\begin{equation}
  \label{eq:outpusonbasis}
  H_k\ket{\bm{\sigma}_j} = c_k(\bm{\sigma}_j)\ket{\bm{\sigma}_j^\prime}.
\end{equation}
Notice, that in a symmetry-adapted basis, evaluating
$c_k(\bm{\sigma}_j)$ requires the evaluation of
\cref{eq:matrixelementrepresentatives_lsed}, where the sublattice
coding technique can be applied. Clearly, we have
\begin{equation}
  \label{eq:matmulkron}
  \braket{\bm{\sigma}_i|\bm{\sigma}_j^\prime} =
  \begin{cases}
    1 \text{ if } \ket{\bm{\sigma}_i} = \ket{\bm{\sigma}_j^\prime}, \\
    0 \text{ else.}
  \end{cases}
\end{equation}
For parallelizing the multiplication \cref{eq:outputvector}, we
distribute the coefficients in the basis $\{ \ket{\bm{\sigma}_i} \}$
onto the different MPI processes. This means we have a mapping,
\begin{equation}
  \label{eq:defprocdistro}
  \texttt{proc}: \ket{\bm{\sigma}_i} \rightarrow \{ 1, \dots, n_{\text{procs}} \},
\end{equation}
that assigns to every basis state of the Hilbert space its MPI process
number. Here, $n_{\text{procs}}$ denotes the number of MPI processes.
In general, $\ket{\bm{\sigma}_j}$ and $\ket{\bm{\sigma}_j^\prime}$ are
not stored in the same process. Hence, the coefficient
$c_k(\bm{\sigma}_j)\braket{\bm{\sigma}_j| \psi_{\text{in}}}$ has to be
sent from the process no. $\texttt{proc}(\ket{\bm{\sigma}_j})$ to
process no. $\texttt{proc}(\ket{\bm{\sigma}_j^\prime})$. This makes
communication between the processes necessary. This communication is
buffered in our implementation, i.e. for every basis state
$\ket{\bm{\sigma}_j}$ we first store the target basis state
$\ket{\bm{\sigma}_j^\prime}$ and the coefficient
$c_k(\bm{\sigma}_j)\braket{\bm{\sigma}_j| \psi_{\text{in}}}$
locally. Once every local basis state has been evaluated, we perform
the communication and exchange the information amongst all
processes. This corresponds to an \texttt{MPI\_Alltoallv} call in the
MPI standard.

After this communication step, every process has to add the received
coefficient to the locally stored coefficient
$\braket{\bm{\sigma}_j^\prime| \psi_{\text{out}}}$. For this, we have
to search, where the now locally stored coefficient of the basis state
$\ket{\bm{\sigma}_j^\prime}$ is located in memory. Typically, we keep
a list of all locally stored basis states defining the position of the
coefficients. This list is then searched for the entry
$\ket{\bm{\sigma}_j^\prime}$, which can also be time-consuming and
needs to be done efficiently. We are thus facing the following
challenges when distributing the basis states of the Hilbert space
amongst the MPI processes:
\begin{itemize}
\item Every process has to know which process any basis state
  $\ket{\bm{\sigma}_i}$ belongs to.
\item The storage of the information about the distribution should be
  memory efficient.
\item The distribution of basis states has to be fair, in the sense
  that every process has a comparable workload in every matrix-vector
  multiplication.
\item The search for a basis state within a process should be done
  efficiently.
\end{itemize}
We now propose a method to address these issues in a satisfactory way.

\paragraph*{Distribution of basis states}
The central point of our parallelization strategy is the proper choice
of the distribution function $\texttt{proc}(\bm{\sigma})$ for the
basis states in \cref{eq:defprocdistro}. We split up every basis state
into prefix and postfix sites,
\begin{equation}
  \ket{\bm{\sigma}} =
  \ket{\underbrace{\sigma_1 \cdots \sigma_{n_{\text{prefix}}}}_{\text{prefix sites}}
    \quad
    \underbrace{\sigma_{n_{\text{prefix}}+1}\cdots\sigma_{n_{\text{prefix}}+ n_{\text{postfix}}}}_{\text{postfix sites}}},
\end{equation}
where $n_{\text{prefix}}$ and $n_{\text{postfix}}$ denote the number
of prefix and postfix sites. We decide that states with the same
prefix are stored in the same MPI process. The prefixes are randomly
distributed amongst all the processes. We do this by using a hash
function that maps the prefix bits onto a random but deterministic MPI
process. This hash function can be chosen such that every process has
a comparable amount of states stored locally. Moreover, a random
distribution of states reduces load balance problems significantly
since the communication structure is randomized. This is in stark
contrast to distributing the basis states in a linear
fashion. Thereby, single processes can often have a multiple of the
workload than other processes, thus causing idle time in other
processes.

\begin{figure}
  \begin{algorithm}[H]
    \caption{Preparation of the distributed and symmetrized Hilbert
      space}
    \begin{algorithmic}[0]
      \State Perform the following steps on every process in parallel
      (no communication necessary) \State \texttt{myid} denotes the
      number of the current MPI process \State prepares data
      structures \texttt{Basis}, \texttt{Limits}
      on each process\\
      \ForEach{prefix spin configuration
        $\ket{\bm{\sigma}_{\text{prefix}}} = \ket{\sigma_1 \cdots
          \sigma_{n_{\text{prefix}}}}$}
      \If{$\texttt{proc}(\ket{\bm{\sigma}_{\text{prefix}}}) \neq $
        \texttt{myid}} \State continue \Else \State \texttt{begin} =
      length(\texttt{Basis}) \ForEach{spin configuration
        $\ket{\bm{\sigma}}$ with prefix
        $\ket{\bm{\sigma}_{\text{prefix}}}$} \State compute
      representative $\ket{\tilde{\bm{\sigma}}}$ of
      $\ket{\bm{\sigma}}$
      \If{$\ket{\bm{\sigma}} = \ket{\tilde{\bm{\sigma}}}$} \State
      append $\ket{\bm{\sigma}}$ to \texttt{Basis} \EndIf \EndForEach
      \texttt{end} = length(\texttt{Basis}) \If{end $\neq$ begin}
      \State insert ($\ket{\bm{\sigma}_{\text{prefix}}}$, begin, end)
      to \texttt{Limits} \EndIf \EndIf \EndForEach
    \end{algorithmic}
    \label{alg:hilbertspaceprep}
  \end{algorithm}
\end{figure}

\begin{figure}
  \begin{algorithm}[H]
    \caption{Parallel matrix-vector multiply for a non-branching term
      $H_k$}
    \begin{algorithmic}[0]
      \Require input wave function $\ket{\psi_{\text{in}}}$ \Ensure
      matrix-vector product
      $\ket{\psi_{\text{out}}} = H_k\ket{\psi_{\text{in}}}$
      %% \State \texttt{myid} denotes the number of the current
      %% MPI
      %% process
      \State $\rhd$ Preparation and sending step (communication may be
      buffered) \ForEach{basis state $\ket{\bm{\sigma}_j}$ stored
        locally in \texttt{Basis}} \State $\cdot$ apply non-branching
      $H_k$ and use sublattice coding \par technique to compute
      $ c_k(\bm{\sigma}_j)$ and $\ket{\bm{\sigma}_j^\prime}$,
      $$
      H_k\ket{\bm{\sigma}_j} =
      c_k(\bm{\sigma}_j)\ket{\bm{\sigma}_j^\prime}.
      $$
      \State $\cdot$ compute
      $c = c_k(\bm{\sigma}_j)\braket{\bm{\sigma}_j| \psi_{\text{in}}}$
      \State $\cdot$ send the pair $(\ket{{\bm{\sigma}}_j^\prime}, c)$
      to process no. $\texttt{proc}(\ket{{\bm{\sigma}}_j^\prime})$
      \EndForEach\\
      \State $\rhd$ Receiving and search step \ForEach{pair
        $(\ket{{\bm{\sigma}}_j^\prime}, c)$ received} \State $\cdot$
      determine indices (\texttt{begin}, \texttt{end}) from
      $\texttt{Limits}(\ket{{\bm{\sigma}}_j^\prime})$ \State $\cdot$
      determine index $i$ of $\ket{\bm{\sigma}^\prime}$ by binary
      search in array \par \texttt{Basis} between (\texttt{begin},
      \texttt{end}) \State $\cdot$ Set
      $\braket{\bm{\sigma}_j^\prime|\psi_{\text{out}}}[i] \leftarrow
      c$ \EndForEach
    \end{algorithmic}
    \label{alg:matrixvector}
  \end{algorithm}
\end{figure}

\begin{figure*}[t]
  \centering
  \begin{minipage}[c]{0.24\textwidth}
    \centering
    \includegraphics[width=0.7\textwidth]{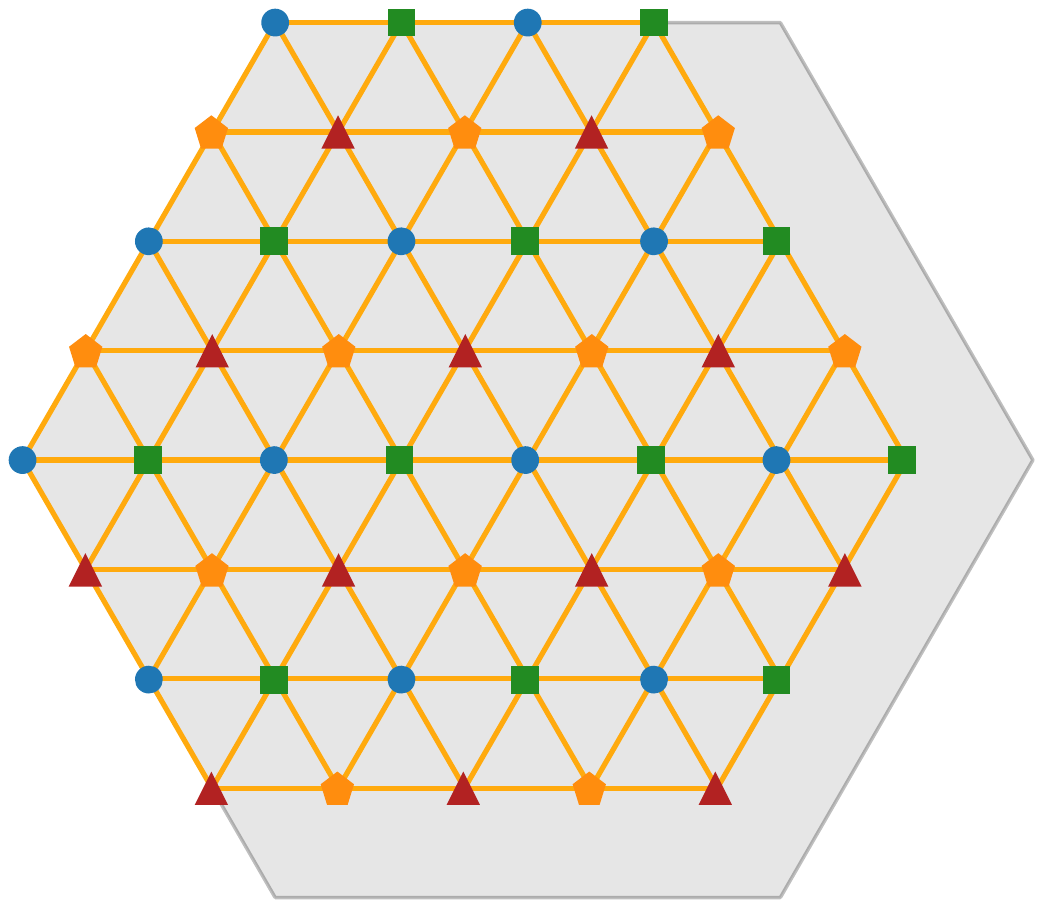}
  \end{minipage}
  \begin{minipage}[c]{0.24\textwidth}
    \centering \includegraphics[width=0.8\textwidth]{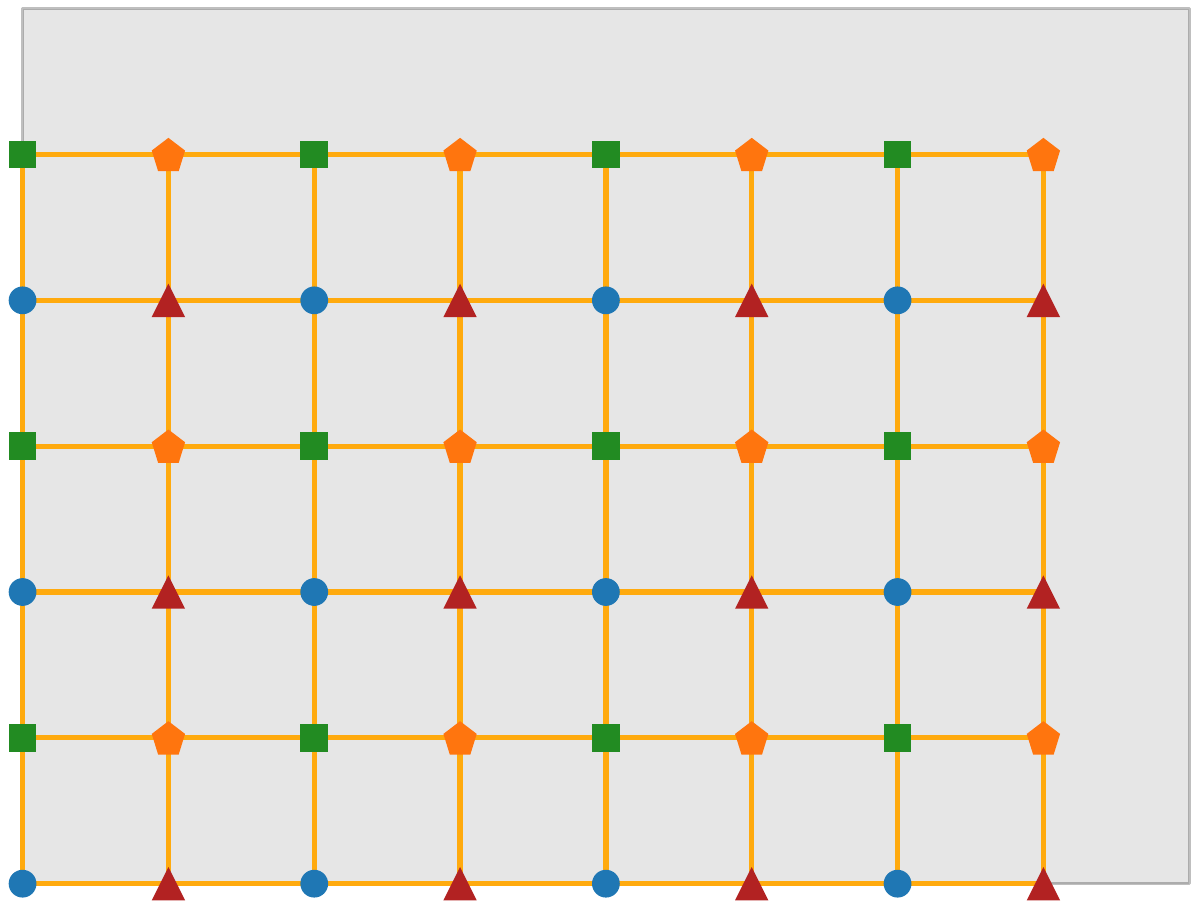}
  \end{minipage}
  \begin{minipage}[c]{0.24\textwidth}
    \centering \includegraphics[width=0.65\textwidth]{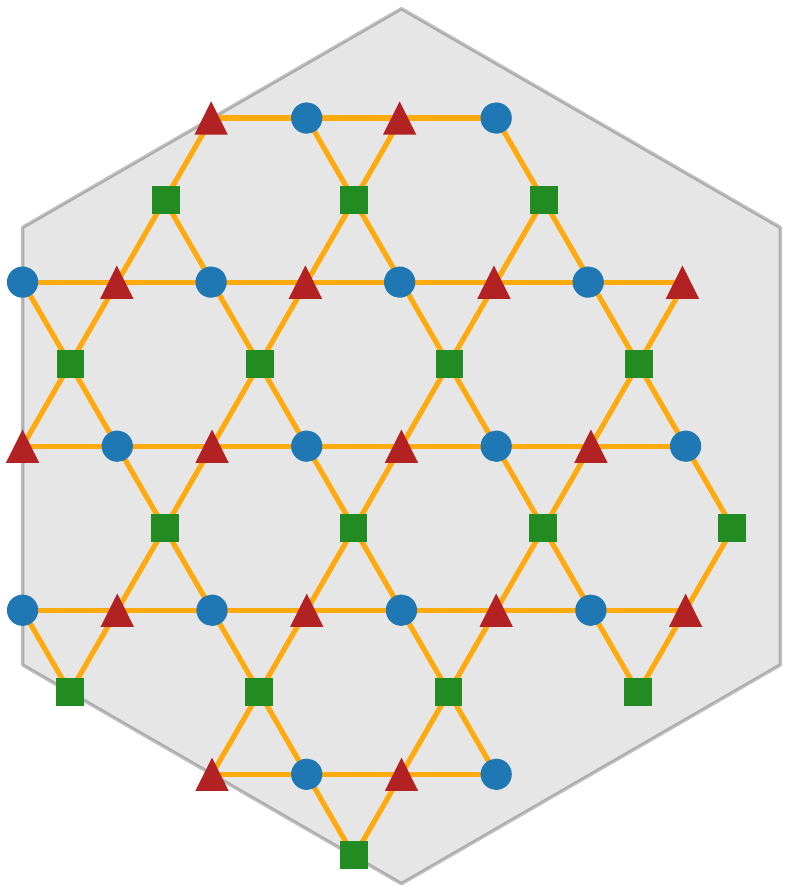}
  \end{minipage}
  \begin{minipage}[c]{0.24\textwidth}
    \centering \includegraphics[width=0.7\textwidth]{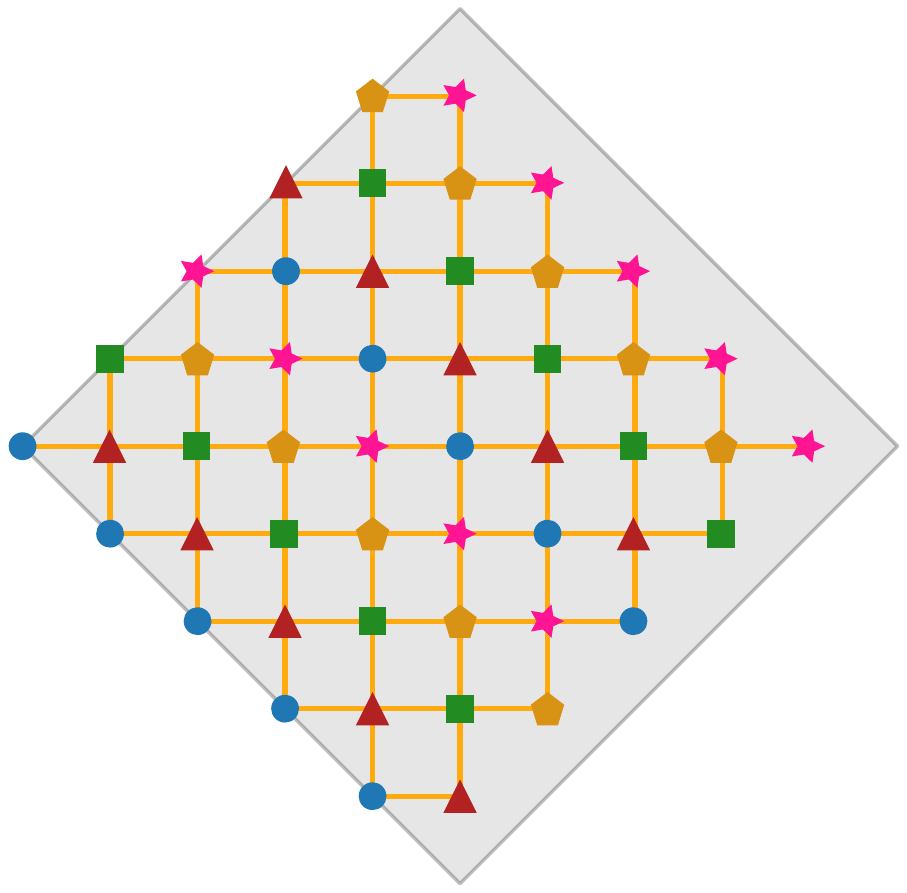}
  \end{minipage}
  \caption{Geometries of Heisenberg spin-$1/2$ model
    benchmarks. Different colors (symbols) show the sublattice
    structure used for the sublattice coding technique. Grey
    background shows the Wigner-Seitz cell defining the periodicity of
    the lattice. \textit{Left}: Triangular lattice, $48$ sites, four
    sublattice structure. \textit{Middle left}: Square lattice, $48$
    sites, four sublattice structure. \textit{Middle right}: Kagome
    lattice, $48$ sites, three sublattice structure. \textit{Right}:
    Square lattice, $50$ sites, five sublattice structure.}
  \label{fig:benchgeo}
\end{figure*}

By choosing this kind of random distribution of basis states, we also
don't have to store any information about their distribution. This
information is all encoded in the hash function.  Nevertheless, we
store the basis states belonging to a process locally in an
array. Finding the index of a given basis state also requires some
computational effort. Here, we use the separation between prefix and
postfix sites.  We store the basis states in an ordered way. This way,
states belonging to the same prefix are aligned in memory as shown in
\cref{fig:edparallelization}. We can store the index of the first and
the last states that belong to a given prefix. To find the index of a
given state we can now lookup the first and last index of the prefix
of this state and perform a binary search for the state between these
two indices. This reduces the length of the array we have to perform
the binary search on and, hence, reduces the computational effort in
finding the index. For implementing this procedure we need two data
structures locally stored on each process.
\begin{enumerate}
\item An array $\texttt{Basis}(i)$ storing all the basis states,
  \begin{equation}
    \texttt{Basis}(i) = \ket{\bm{\sigma}_i}, \quad i=1,\ldots,D.
  \end{equation}
\item An associative array
  $\texttt{Limits}(\ket{\bm{\sigma}_{\text{prefix}}})$ storing the map
  \begin{equation}
    \texttt{Limits}(\ket{\bm{\sigma}_{\text{prefix}}}) =
    \left[\texttt{begin}(\ket{\bm{\sigma}_{\text{prefix}}}),
      \texttt{end}(\ket{\bm{\sigma}_{\text{prefix}}})\right]
  \end{equation}
  where $\texttt{begin}(\ket{\bm{\sigma}_{\text{prefix}}})$ denotes
  the index of the first state with prefix
  $\ket{\bm{\sigma}_{\text{prefix}}}$ and
  $\texttt{end}(\ket{\bm{\sigma}_{\text{prefix}}})$ denotes the index
  of the last state with this prefix in the array $\texttt{Basis}(i)$,
  $\ket{\bm{\sigma}_{\text{prefix}}} = \ket{\sigma_1 \cdots
    \sigma_{n_{\text{prefix}}}}$.
\end{enumerate}
In \cref{alg:hilbertspaceprep} we summarize how to prepare these data
structures. The parallel matrix-vector multiplication in pseudocode is
shown in \cref{alg:matrixvector}. When working in the symmetry-adapted
basis, the lookup tables of the sublattice coding method need to be
accessible to every MPI process. One way to achieve this is of course,
that every process generates its own lookup tables.  However, in
present-day supercomputers, several processes will be assigned to the
same physical machine sharing the same physical memory.  To save
memory, the lookup tables are stored only once on a computing
node. Its processes can then access the lookup tables via shared
memory access. In our code, we use POSIX shared memory
functions~\cite{posixshm} to implement this hybrid parallelization.

\section{Benchmarks}
In order to assess the power of the methods proposed in the previous
sections, we performed test runs to compute ground state energies. We
considered the Heisenberg antiferromagnetic spin-$1/2$ nearest
neighbor model,
\begin{equation}
  H = J\sum\limits_{\langle i,j \rangle}\bm{S}_i\cdot\bm{S}_j,\quad J=1,
  \label{eq:heisenbergmodellattice}
\end{equation}
on four different lattice geometries: square ($48$ sites), triangular
($48$ sites), kagome ($48$ sites) and square ($50$
sites). \Cref{fig:benchgeo} shows the simulation clusters and
the sublattice structure we used. The benchmarks were performed on
three different supercomputers. The Vienna Scientific Cluster VSC3 is
built up from over 2020 nodes with two Intel Xeon E5-2650v2, 2.6 GHz,
8 core processors, the supercomputer Hydra at the Max Planck
Supercomputing \& Data Facility in Garching with over 3500 nodes with
20 core Intel Ivy Bridge 2.8 GHz processors and the System B Sekirei
at the Institute for Solid State Physics of the University of Tokyo
with over 1584 nodes with two Intel Xeon E5-2680v3 12 core 2.5GHz
processors. Both the Hydra and Sekirei use InfiniBand FDR
interconnect, whereas the VSC3 uses Intel TrueScale Infiniband for
network communication.

The benchmarks are summarized in \cref{tab:edbenchmarks}. We make use
of all translational, certain point group symmetries and spin-flip
symmetry. We show the memory occupied by a single lookup table for the
symmetries. Since we use a single buffered and blocking all-to-all
communication in the implementation it is straightforward to measure
the percentage of time spent for MPI communication by taking the time
before the communication call and afterward. In order to validate the
results of our computation, we compared the results of the
unfrustrated square case to Quantum Monte Carlo computations of the
ground state energy. We used a continuous time world-line Monte Carlo
Code \cite{Todo2001a} with $10^5$ thermalization and $10^6$
measurements at temperature $T=0.01$.  The computed energies per site
are $E/N = -0.676013 \pm 2\cdot 10^{-5}$ for the $48$ site square
cluster and $E/N = -0.67512 \pm 2\cdot 10^{-5}$ for the $50$ site
cluster. The actual values computed with ED are within the error bars.
% \begin{itemize}
% \item 48 QMC -0.676013 +- 2.02829e-05 100000 therm 1000000
%   production
% \item 50 QMC -0.67512 +- 1.98685e-05 100000 therm 1000000 production
% \end{itemize}
The ground state energy of the kagome Heisenberg antiferromagnet on
$48$ sites has been previously computed \cite{Lauchli2016} with a
specialized code and agrees with our results. We see that the amount
of time spent for communication is different for the three
supercomputers. On Sekirei, a parallel efficiency of $61\%$ on 3456
cores has been achieved. 

\begin{figure}[h]
  \centering
  \includegraphics[width=0.5\textwidth]{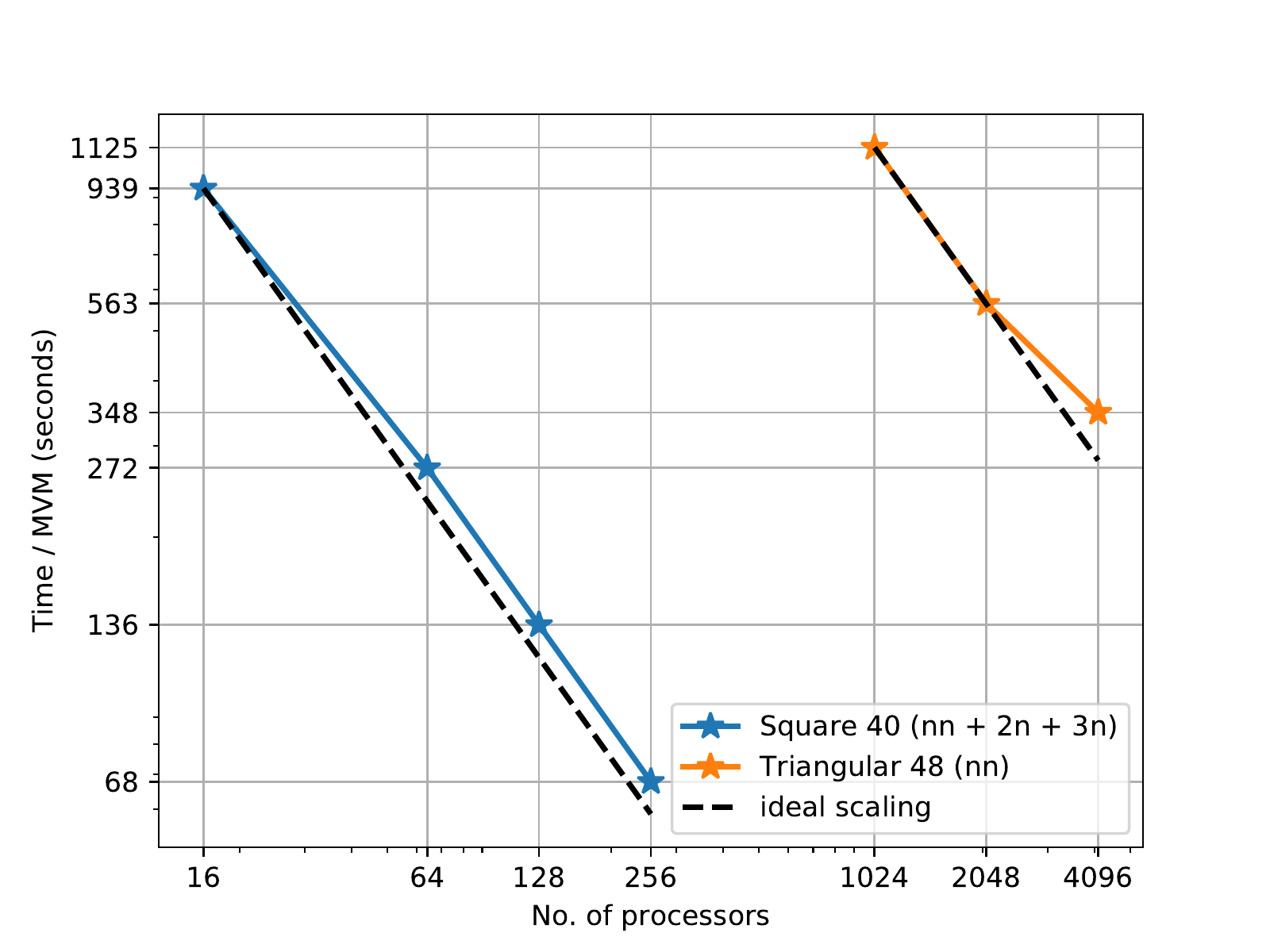}
  \caption{Scaling behaviour of the parallelization. The time for one matrix-vector 
    multiplication in seconds is compared for the total number of processes.
    Both axes are scaled logarithmically. The matrix is the Hamiltonian of the 
    Heisenberg spin $1/2$ antiferromagnet on a $40$ site cluster 
    with additional next-nearest neighbour and third-nearest neighbour interactions for the  
    benchmark on $16$, $64$, $128$ and $256$ cores. For $1024$, $2048$ and $4096$ cores
    the matrix is the Hamiltonian of the Heisenberg spin $1/2$ antiferromagnet on a $48$ site
    triangular cluster. The Hamiltonian is considered in its ground state sector. 
    We observe almost ideal scaling behaviour up to $4096$ processors.    
  }
  \label{fig:strongscaling}
\end{figure}

Results for running the same problem on various numbers of processors 
are shown in \cref{fig:strongscaling}. We chose two different problems for 
two sets of number of cores, the Heisenberg antiferromagnet with additional 
next-nearest neighbour and third nearest neighbour interactions on a $40$ 
site square lattice and the Heisenberg antiferromagnet on a $48$ site triangular lattice. 
We observe almost ideal scaling behaviour up to $4096$ MPI processes. Hence, 
the parallelization strategy described above successfully solves load balancing 
problems in the MPI communication. This benchmark has been performed on the 
Curie supercomputer at GENCI-TGCC-CEA, France. 

\begin{table*}[t]
  \centering
  \begin{tabular}{|c | c | c | c | c |}
    \hline
    Geometry & Triangular 48 & Square 48 & Kagome 48 & Square 50 \\
    computer & Sekirei & VSC3 & Hydra & Sekirei \\
    \hline
    point group & D6 & D2 & D6 & D2 \\
    No. of symmetries & 1152 & 384 & 384 & 400 \\
    % \# sublattices & 4 & 4 & 3 & 5 \\
    dimension & $2.8 \cdot 10 ^{10}$ &  $8.3 \cdot 10^{10}$ & $8.4 \cdot 10^{10}$ & $3.2 \cdot 10^{11}$ \\
    No. of cores & 3456 & 8192 & 10240 & 3456 \\
    total memory &  2.5 TB & n.A. & n.A. & 15.5 TB \\
    memory lookup & 151 MB & 50 MB & 604 MB & 17 MB \\
    Time / MVM & 399 s & 1241 s & 258 s & 3304 s \\
    \% comm. time & 39\% & 77\% & 48\% & 39\% \\ \hline
    g.s. sector & $\Gamma$.A1.even & $\Gamma$.A1.even & $\Gamma$.A1.even & M.A1.odd\\
    g.s. energy & -26.8129452715 & -32.4473598728 & -21.0577870635 & -33.7551019315\\
    \hline
  \end{tabular}
  \caption{Benchmark results for various problems on three different
    supercomputer systems described in the main text. The employed
    symmetries include translational, point group and spinflip
    symmetry. We show the total memory used by all MPI processes and
    the memory used by the lookup tables for the sublattice coding
    technique. We also show the amount of time spent for
    communication. For labeling the ground state representations
    $\Gamma$ denotes the $(0,0)$ and $M$ the $(\pi,\pi)$ point in the
    Brillouin zone. A1 denotes the trivial point group representation
    and even/odd denotes the spinflip symmetry representation.}
  \label{tab:edbenchmarks}
\end{table*}

\section{Discussion}
\label{sec:eddiscussion}
The sublattice coding technique presented in
\cref{sec:sublatticecodingtechnique} allows for fast and
memory-efficient evaluation of the matrix elements in a
symmetry-adapted basis
\cref{eq:matrixelementrepresentatives_lsed}. Still, the construction
of the sublattices imposes restrictions on the geometry of the
simulation cluster. A sublattice construction with $N_{\text{sublat}}$
sublattices at least requires the number of sites to be divisible by
$N_{\text{sublat}}$. The sublattice coding technique yields no
advantages for lattice samples that have a prime number of sites. For
lattices with several basis sites per unit cell a natural sublattice
decomposition exists. The sublattices are given by the lattices
defined by the corresponding basis sites. This is the case for the
honeycomb and kagome lattice, where a natural two (resp. three)
sublattice decomposition exists, cf.~\cref{fig:sublatticeorderings}.

The sublattice decomposition for a given lattice is not unique. This
can be seen in the case of the $50$ site square lattice, whose five
sublattice decomposition is shown in \cref{fig:benchgeo}. As
a bipartite square lattice, it also allows for a two-sublattice
decomposition. Three- and four-sublattice decompositions exist for
other square lattice clusters as well (see
e.g. \cref{fig:benchgeo}). Hence, for a given simulation
cluster there may exist more than one sublattice decomposition. Most
of the high symmetry square and triangular lattice samples possess at
least one sublattice decomposition in two or more sublattices.  Still,
a given sublattice decomposition may restrict the symmetry group if
certain symmetry elements split up sublattices. This is, for example,
the case for the $50$ site square lattice in
\cref{fig:benchgeo}. While the cluster itself has a full
four-fold rotational and reflectional symmetry, the $90^\circ$
rotation is not sublattice stable. Therefore, only the $180^\circ$
rotation and reflection symmetry has been used as point group
symmteries in the computation.

Increasing the number of sublattices decreases the memory required for
storing the lookup tables. The computational effort for computing the
representative as in \cref{alg:sublatticecodingfindrep} increases
linearly in the number of sublattices. Also, smaller sublattices yield
more potential representative symmetries \cref{eq:repsymdef} that have
to be applied to the spin configuration. In principle, there is no
restriction on the number of sublattices and the proper choice depends
on the geometry of the simulation cluster, the available memory, and
the desired speed. Fewer sublattices allow
\cref{alg:sublatticecodingfindrep} to evaluate matrix elements faster.

The method of distributing the basis states of the Hilbert space is
independent of the sublattice coding algorithm. Hence, this kind of
parallelization can also be applied to problems without symmetries,
like disordered systems. All information about the distribution of
basis states is encoded by the hash function, that can be of rather
simple type to achieve a balanced distribution.

One main motivation for performing large-scale ED computations is
reaching system sizes for which high symmetry clusters are available.
The possibility to simulate $48$ spin-$1/2$ particles gives access to
the interesting triangular and kagome lattices shown in
\cref{fig:benchgeo}. These samples
both have full sixfold rotational and reflectional symmetry. In
reciprocal space, these clusters both accommodate the $K$ point and
even the $M$ point for the triangular case. This feature is important
to distinguish different phases with different ordering vectors. For
the square lattice case, an interesting $52$ site cluster with
four-fold rotational symmetry exists featuring the $(\pi,\pi)$ and
$(\pi,0)$ point in reciprocal space. A study of the Heisenberg model
with next-nearest neighbor interactions on this cluster can,
therefore, yield valuable insights into nature of the intermediate
phase, whose nature is not fully understood as of today. The methods
proposed in this manuscript allow for these calculations on large
present-day supercomputers, since the Hilbert space dimension of this
problem is roughly four times larger than the investigated $50$ site
case and a four sublattice decomposition is available.

Apart from spin systems, the sublattice coding algorithm also applies
to fermionic systems, when the Hamiltonian is expressed in the occupation
number basis. The occupation numbers of the orbitals on the respective 
sublattices define the sublattice configurations. In addition to 
computing a representative and representative symmetry, a Fermi sign 
has to be computed to evaluate matrix elements, which can be done 
efficiently. 

\section{Conclusion}
\label{sec:edconclusion}
We proposed the generic sublattice coding algorithm for making
efficient use of discrete symmetries in large-scale ED
computations. The method can be used flexibly on most lattice
geometries and only requires a reasonable amount of memory for storing
the lookup tables. The parallelization strategy for distributed memory
architectures we discussed includes a random distribution of the
Hilbert space amongst the parallel processes. Lookup tables of the
sublattice coding technique are stored only once per node and are
accessed via shared memory. Using these techniques, we showed that
computations of spin-$1/2$ models of up to 50 spins have now become
feasible.

\section*{Acknowledgements}
The ED calculations of the $50$ site cluster have been performed using
the facilities of the Supercomputer Center, the Institute for Solid
State Physics, the University of Tokyo. The scaling benchmarks in 
\cref{fig:strongscaling} have been performed on the supercomputer Curie 
(GENCI-TGCC-CEA, France).
We especially thank Synge Todo, Roderich Moessner and Sylvain Capponi for 
making some of these simulations possible. A.W. acknowledges support 
through the Austrian Science Fund project I-1310-N27 (DFG FOR1807) 
and the Marietta Blau-Stipendium of OeAD-GmbH, financed by the Austrian 
Bundesministeriums für Wissenschaft, Forschung und Wirtschaft (BMWFW). 
Further computations for this manuscript have been carried out on VSC3 
of the Vienna Scientific Cluster, the supercomputer Hydra at the Max Planck
Supercomputing \& Data Facility in Garching.

\bibliography{short_journal_names.bib,ed_paper.bib}

\end{document}